\documentclass{aa}  

\usepackage{graphicx}
\usepackage{txfonts}
\begin{document}

   \title{Understanding the environment around the intermediate mass black hole candidate ESO 243-49 HLX-1}


   \author{N. A. Webb
          \inst{1,2}
          \and
          A. Gu\'erou\inst{1,3}
          \and
          B. Ciambur\inst{4}
          \and
          A. Detoeuf\inst{1,2}
          \and
          M. Coriat\inst{1,2}
          \and
          O. Godet\inst{1,2}
          \and
          D. Barret\inst{1,2} \and F. Combes\inst{5} \and T. Contini\inst{1,2} \and Alister W. Graham\inst{4} \and T. J. Maccarone\inst{6} \and M. Mrkalj\inst{1,2} \and M. Servillat\inst{7}   \and I. Schroetter\inst{1,2} \and K. Wiersema\inst{8}
          }

   \institute{Universit\'e de Toulouse; UPS-OMP, IRAP,  Toulouse, France\\
              \email{Natalie.Webb@irap.omp.eu}
         \and
             CNRS, IRAP, 9 av. Colonel Roche, BP 44346, F-31028 Toulouse cedex 4,  France
\and
European Southern Observatory, Karl-Schwarzschild-Str. 2, D-85748 Garching, Germany
          \and
Centre for Astrophysics and Supercomputing, Swinburne University of Technology, Hawthorn, VIC 3122, Australia
\and
Observatoire de Paris, LERMA, College de France, CNRS, PSL Univ., Sorbonne Univ.
UPMC, F-75014, Paris, France
\and
Department of Physics, Texas Tech University, Box 41051, Lubbock, TX 79409-1051, USA
\and
LUTH, Laboratoire Univers et Th\'eories (CNRS/INSU, Observatoire de Paris, Universit\'e Paris Diderot), 5 place Jules Janssen, F-92190 Meudon, France
\and
Department of Physics \& Astronomy, University of Leicester, Leicester, LE1 7RH, UK
             }

   \date{Received ; accepted }

 
  \abstract
   {}
   {ESO 243-49 HLX-1, otherwise known as HLX-1, is an intermediate mass black hole (IMBH) candidate located 8\arcsec\ (3.7 Kpc) from the centre of the edge-on S0 galaxy ESO 243-49. How the black hole came to be associated with this galaxy, and the nature of the environment in which it resides, are still unclear. Using multi-wavelength observations we investigate the nature of the medium surrounding HLX-1, search for evidence of past mergers with ESO 243-49 and constrain parameters of the galaxy, including the mass of the expected central supermassive black hole, essential for future modelling of the interaction of the IMBH and ESO 243-49.}
   {We reduce and analyse integral field unit observations of ESO 243-49 that were taken with the {\em MUSE} instrument on the VLT. Using complementary multi-wavelength data, including {\em X-Shooter}, {\em HST}, {\em Swift}, {\em Chandra} and {\em ATCA} data, we further examine the vicinity of HLX-1. We additionally examine the nature of the host galaxy and estimate the mass of the central supermassive black hole in ESO 243-49 using (black hole mass)--(host spheroid) scaling relations and the fundamental plane of black hole activity.}
   {No evidence for a recent minor-merger that could result in the presence of the IMBH is discerned, but the data are compatible with a scenario in which  minor mergers may have occurred in the history of ESO 243-49. The {\em MUSE} data reveal a rapidly rotating disc in the centre of the galaxy, around the supermassive black hole. The mass of the supermassive black hole at the centre of ESO 243-49 is estimated to be 0.5-23 $\times$ 10$^7$ M$_\odot$.  Studying the spectra of HLX-1, that were taken in the low/hard state, we determine H$_\alpha$ flux variability to be at least a factor 6, compared to observations taken during the high/soft state. This H$_\alpha$ flux variability over one year indicates that the line originates close to the intermediate mass black hole, excluding the possibility that the line emanates from a surrounding nebula or a star cluster. The large variability associated with the X-ray states of HLX-1 confirms that the H$_\alpha$ line is associated with the object and therefore validates the distance to HLX-1.}
   {}


   \maketitle
%

\section{Introduction}
\label{sec:intro}

The galaxy \object{ESO 243-49} situated at $z=0.023$, is an almost edge-on, S0 or early-type spiral, located 0.3 Mpc from the central dominant galaxy in the cluster \object{Abell 2877} \citep{sant08}. \object{ESO 243-49} is one of the larger galaxies in the cluster \citep{huds01}.  Given its size and proximity to the host cluster centre, \object{ESO 243-49} is expected to have suffered dynamical effects, such as interactions or accretion of other bodies, though little evidence for such events were noted by  \cite{sant08} after analysing $R$-band data of this galaxy.

\object{ESO 243-49} is however fairly exceptional as it hosts the intermediate mass black hole (IMBH, $\sim$10$^{2-5}$ M$_{\odot}$) candidate \object{ESO 243-49 HLX-1} \citep{farr09}.  The observational evidence for the existence of IMBHs has until recently been weak, with few convincing examples, but they have been proposed to play a role in a number of astrophysical scenarios, including the formation of supermassive black holes through mergers and accretion \citep[e.g.][]{mada01}. Some IMBHs may avoid such mergers or may even be ejected during merger interactions \citep{merr04}. Studying these IMBHs and their environment should allow us to understand their origin, either as remnant primordial BHs \citep[][e.g.]{yoko97,duec04}, or from runaway stellar collisions \citep[e.g.][]{gier15,arca16}, or as the remnants of population III stars \citep[e.g.][]{rico16} for example and how they are fuelled \citep[from stellar winds, via Bondi accretion, or from Roche lobe overflow, e.g.][]{mill02,mill04}. 

HLX-1 is situated just 8$\arcsec$ from the centre of \object{ESO 243-49}. HLX-1 was shown to be associated with the galaxy using optical spectroscopy of the optical counterpart \citep{sori10} which revealed an emission line consistent with the Balmer H$_\alpha$ line compatible with the distance of \object{ESO 243-49} \citep{wier10,sori13}. The physical process resulting in an  H$_\alpha$ emission line is unclear, where it could be from an accretion disc around the IMBH, created by material accreted from a companion star \citep{laso11} or the host star cluster or a photo-ionised/shock-ionised gas nebula close to HLX-1 \citep{wier10}. However, as \cite{sori13} stated, the  narrow  FWHM  of  the  line  observed from HLX-1 either requires an extremely face-on disk ($\leq$3$^\circ$) or, implies that the line does not originate from a Keplerian disc. They suggest that it may then come from a low density nebula around the IMBH. 

\begin{figure*}
\includegraphics[width=18cm]{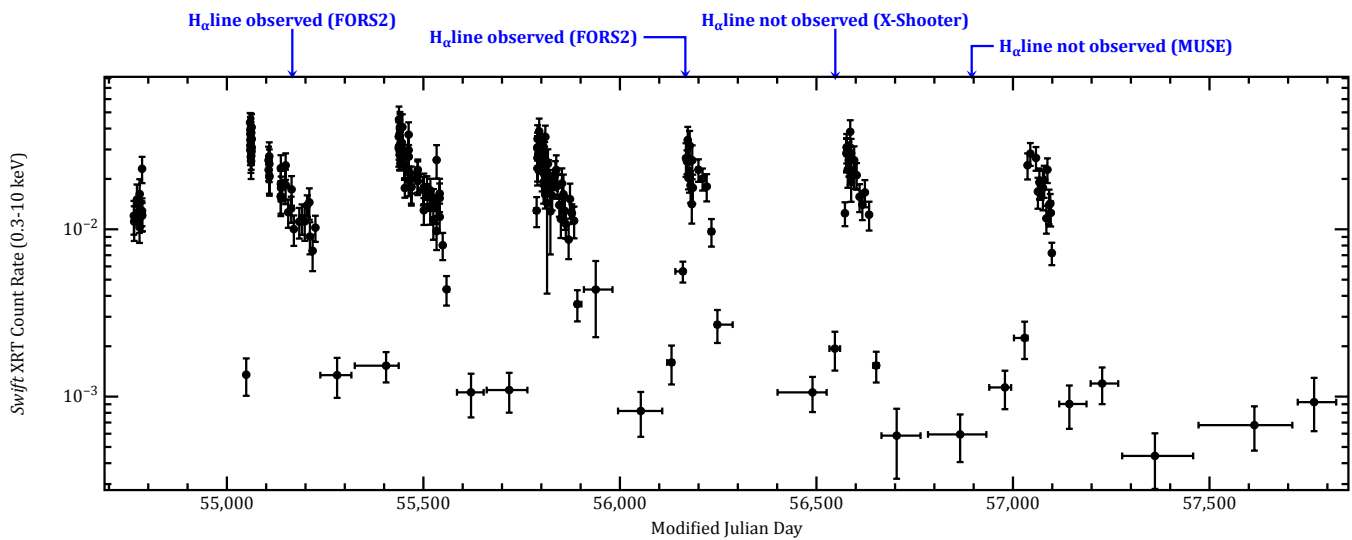}
\caption{Swift X-ray lightcurve of HLX-1 from 2008-2016. The four arrows show times when optical spectroscopy was taken and whether or not a significant detection of the H$_\alpha$ line was made.}
\label{fig:XrayLC}
\end{figure*}

The X-ray flux from HLX-1 has been shown to vary by a factor of $\sim$50 \citep[][and Fig.~\ref{fig:XrayLC}]{gode12}, quasi-periodically, initially with a period close to one year \citep{laso11}, but more recently, the period between outbursts has lengthened significantly, which may be due to the companion star becoming unbound from the IMBH \citep{gode14}. Whilst HLX-1 is faint, the X-ray emission is hard, typical of the X-ray emission seen in X-ray binaries in the low/hard state \citep{gode09a}. In the bright state, the emission is soft \citep{gode09a,serv11,yan15}, typical of the X-ray emission seen in X-ray binaries in the high/soft state.  Moreover, radio flares emitted during the transition from the low/hard to the high/soft state have also been observed \citep{webb12}, adding to the similarities between HLX-1's behaviour and that of X-ray binaries. The increase of flux in the case of HLX-1 was thought to be due to matter being tidally stripped from the companion as it passes at periastron, falling onto the IMBH \citep{laso11}, but this model has recently been questioned by various authors \citep[e.g.][]{king14,laso15}.  

The mass of HLX-1 has been estimated to be $\sim$10$^4$ M$_\odot$ from modelling the X-ray spectrum with a variety of accretion disc models and from Eddington scaling \citep{gode12,serv11,davi11,webb12,stra14}.  The formation mechanism for HLX-1 is not clear from previous observations. \cite{webb10} proposed that HLX-1 could be an intermediate
mass black hole that was once at the centre of a dwarf galaxy, that had interacted with the host galaxy \object{ESO 243-49}, possibly like LEDA 87300 \citep{bald15,grah16}.  \cite{farr12} suggested that this merger may have been recent, possibly accounting for the presence of dusty lanes in the central regions of \object{ESO 243-49}. However, their study using {\em Swift} X-ray data and {\em HST} ultra-violet to infra-red photometry could not exclude that HLX-1 resides in an old stellar population, reminiscent of a globular cluster, one of the places IMBH are expected to form \citep[e.g.][]{mill04}. 

To test the merger scenario, \cite{musa15} took {\em Australia Telescope Compact Array} (ATCA) H{\sc I} radio observations of ESO 243-49 and the surrounding neighbourhood, in order to search for signatures of a recent merger event.  No H{\sc I} emission was detected, indicating that any recent merger that may have taken place was not gas rich. The absence of gas may be due to the cluster environment depleting ESO 243-49's H{\sc I} gas reservoir.  They also identify another galaxy in the cluster \object{Abell 2877} undergoing a merger, where gas is being depleted due to the cluster.

\cite{mape12} investigated the minor merger scenario using N-body/smoothed particle hydrodynamics simulations and concluded that it was indeed possible.  However, to make realistic simulations of a merger between ESO 243-49 and another body that would give rise to the IMBH HLX-1, it is essential to know the mass of the supermassive black hole (SMBH) at the core of \object{ESO 243-49} and the mass of the galaxy.  The mass of SMBH can be estimated through the stellar velocity distribution of the central regions of the galaxy \citep[e.g.][]{gult09,grah13}. Alternatively, the mass can be estimated thanks to the correlation with the bulge luminosity. Such single parameter scaling relations, and indeed why we came to believe in massive BHs, are extensively reviewed in Graham (2016).   A third independent method for estimating the black hole mass is the {\em black hole fundamental plane} \citep[e.g.][]{merl03,koer06}. This plane best relates the radio emission for black holes in the low/hard state, the X-ray emission and the mass across more than eight orders of magnitude in black hole mass.

The stellar kinematics and population (i.e., age and metallicity), as well as the gas content are further parameters that can constrain realistic merger simulations. To make these constraints, we observed the galaxy \object{ESO 243-49} using the new integral field unit on the Very Large Telescope (VLT), {\em Multi Unit Spectroscopic Explorer} \citep[MUSE,][]{baco10}.  The data were taken in August 2014 during the Science Verification time for this new instrument. These data are the subject of the current paper, combined with {\em X-Shooter} data of \object{HLX-1} to make constraints on the nature of this source. We also re-analyse some archival data to estimate the mass of the central supermassive black hole in \object{ESO 243-49} using the different methods described above.



\section{Data reduction}

A brief overview of the observations used in this paper are given in Table~\ref{tab:obs}. The details of each data set are given in the following subsections.

\subsection{MUSE data}

Twelve 500s observations centred on \object{ESO 243-49} (RA = 01$^h$10$^m$27$\fs$746, dec= -46$^\circ$04$^\prime$27$\farcs$41, J2000) were taken with {\em MUSE} installed on UT4 of the VLT (PI: N. Webb). Four observations were taken on 2014 August 26, along with one blank sky image (taken after two target observations) and the remaining 8 were taken the following night, with a further two blank sky observations, one taken after the first pair of observations and a second after the third pair.  All the observations were made whilst HLX-1 was in the low/hard state, as determined from quasi-simultaneous {\em Swift} X-ray data (see Fig.~\ref{fig:XrayLC}). Bias, dark, flat and arc observations were taken each night and a field for the astrometric model was taken the 26th. Spectrophotometric standards were taken on the 27th August 2014.  {\em MUSE} was used in the wide field mode, which has a field of view of 1 $\times$ 1 arcmin$^2$ sampled at 0.2\arcsec. The spectral range covers 4750-9300 \AA, with a medium spectral resolution (average $\sim$3000). A maximum of 1\arcsec\ was recorded for the seeing. We used the {\em Recipe Flexible Execution Workbench} \citep[Reflex,][]{hook08} {\em MUSE} Workflow v1.0.1 \citep{weil14} to reduce the data. The data reduction was done in a similar way to \cite{baco15}. Bias, arcs and flat field master calibration solutions were created using the relevant exposures obtained each night. The bias and dark masters were subtracted from the ESO 243-49 observations, the blank sky observations and the sky flats. The observations were then divided by the master flats. An additional flat-field correction was made with the sky flats to correct for the difference between sky and calibration unit illumination. The wavelength calibration was done using the 15 Ne, Xe, HgCd arc observations made on each of the two observing nights. Further geometric and wavelength corrections were carried out using tables from the {\em MUSE} consortium, to calibrate in space and wavelength. The sky continuum and line correction was done by using the task {\em muse\_create\_sky} on observations of blank sky close to ESO 243-49 and taken between two observations of the galaxy and then subtracting these from the target observations. The astrometric fields and spectrophotometric standards were used to find the astrometric solution and calibrate the flux (respectively).  The data cube was produced from pixel-tables using a 3D drizzle interpolation process which included sigma-clipping to reject outliers such as cosmic rays. Using 5 bright, but non-saturated stars in the field, the offset between each observation was calculated using {\em iraf} and then all 12 data cubes were combined to make one single deep exposure.

\begin{table}
\caption{Summary of the observations used in this paper}\label{tab:obs}
\centering
\begin{tabular}{lccc}
\hline \hline
Instrument & Data & Date & MJD  \\
\hline
MUSE/VLT & optical  & 26-27 Aug 2014 & 56895-6\\
 & spectroscopy &  & \\
X-Shooter & UV-IR & 3, 6, & 56538,41, \\
/VLT & spectroscopy & 29 Sep 2013 & 56564\\
WFC3/HST & H-band & 23 Sep 2010 & 55462\\
ATCA & 5.5+9 GHz& 13 Sep 2010& 55452\\
ACIS/Chandra & 0.5-10 keV & 6 Sep 2010 & 55445\\
XRT/Swift  & spectrum & 13-14 Sep 2010 & 55452-3\\
XRT/Swift  & lightcurve & Oct 2008 & 54763-\\
 &  & -Aug 2016 & 57628\\
\hline
\end{tabular}
\end{table}

\subsection{X-Shooter}

A total of twenty exposures was taken with the VLT instrument {\em X-Shooter} \citep{vern11} on the 3, 6 and 29 September 2013 (PI: N. Webb). These were centred on the intermediate mass black hole candidate \object{ESO 243-49 HLX-1} \citep[RA = 01$^h$10$^m$28$\fs$3, dec= -46$^\circ$04$^\prime$22$\farcs$3,][]{webb10}. Two observations were made on 3 September, 8 on 6 September and 10 more on the 29 September, whilst HLX-1 was in the low/hard state. The combined exposures totalled 16360s in the 5595-10240\AA\ (VIS) band, 12000s in the domain 10240-24800\AA\ (NIR) and 17400s in the domain 3000-5595\AA\ (UVB). The slit, of length 11$\arcsec$  was placed at the parallactic angle for all the observations. The slit width was set to 0.9$\arcsec$ for the VIS and NIR observations and 1.0$\arcsec$ for the UVB observations. The spectral resolution was 5100 for UVB, 8800 for VIS and 5300 for NIR for a source such as the galaxy filling the slit, but slightly better for a point like source with good seeing.  The seeing was between 0.8-0.9$\arcsec$ on 3 September, $\sim$0.7$\arcsec$ on 6 September and just over 1$\arcsec$ on 29 September.

The data were reduced with Gasgano and the X-Shooter pipeline \citep{modi10}. A master bias was created for each night of observations for the UVB and VIS observations and subsequently subtracted from the data. A master dark was created and subtracted from the NIR data.  The data were flatfielded and the orders were traced, the spectra were extracted and calibrated in wavelength and flux. The HLX-1 spectrum was derived from the combined spectrum and spectra of the galaxy ESO 243-49, from either side of HLX-1, were also determined.  The flux due to the galaxy at the position of HLX-1 was determined from the adjacent galaxy spectra and this subtracted from the HLX-1 spectrum, to correct for the diffuse galactic emission. 

We also took advantage of the release of {\em X-shooter} science data products. The 2D spectra were downloaded from the ESO archive and then {\em iraf} tools, notably {\em apall} were used to extract the 1D spectra of HLX-1 and the galaxy. The galaxy spectrum, scaled to the value at the position of HLX-1, was subtracted from the HLX-1 spectrum as before. 

\subsection{HST data}

An 806 s exposure in the $F160W$ ($H$-band) of the galaxy \object{ESO 243-49} was obtained with the {\em Wide Field Camera 3} onboard the {\em Hubble Space Telescope} ({\em HST}) on 23 September 2010, as described in \cite{farr12}.  The reduced images were analysed using the new tasks {\em Isofit} and {\em Cmodel} which comprise a new fitting formalism for isophotes to allow more accurate modelling of galaxies with non-elliptical shapes, such as disk galaxies viewed edge-on. These tasks are described in detail in \cite{ciam15}.  We extracted the 1-D major-axis surface brightness profile, $\mu(R_{\rm maj})$, which we further decomposed into its photometric constituents with the software $Profiler$, which solves for the best-fit parameters by minimising the RMS (root mean square) scatter $\Delta_{\rm rms} = \sqrt{\sum_{i=1,n} (\mu_{{\rm data},i} - \mu_{{\rm model},i})^2 /(n-\nu)}$, where $n$ is the number of data points, $\nu$ is the number of degrees of freedom and $\mu_{\rm model}$ is generated at each iteration and consists of the summed components convolved with the PSF \citep{ciam16}. The PSF was estimated by fitting a Moffat profile to 8 bright, unsaturated stars in the image, with the IRAF task $Imexamine$.

\subsection{ATCA data}

ESO 243-49 was observed with the Australia Telescope Compact Array (ATCA) on 13 September 2010 (PI: N. Webb), using the upgraded Compact Array Broadband Backend \cite[CABB,][]{wils11}. The data were taken using the CFB 1M-0.5k correlator configuration with 2 GHz bandwidth and 2048 channels, each with 1 MHz resolution. The observation was performed at the central frequencies of 5.5 GHz and 9 GHz simultaneously. The array was in the 750 m configuration (giving baselines up to 5 km when all 6 antennas are used).  The total on-source integration time was $\sim$11 h. The primary calibrator PKS 1934-638 was used for absolute flux and bandpass calibration, while the secondary calibrator 0048-427 was used for the phase and antenna gain calibration. For each observation, we observed 1934-638 for 10 min and the phase calibrator was observed every 15 min. 

The data reduction and analysis was performed, as described in \cite{webb12}, with the Multichannel Image Reconstruction, Image Analysis and Display (MIRIAD) software \citep{saul95}. 

\subsection{Chandra data}

10 ks of X-ray data were taken with the {\em AXAF CCD Imaging Spectrometer} (ACIS) onboard {\em Chandra} on 2010 September 6 (ObsID 13122, MJD 55445). These data were reduced in the standard way, as described in \cite{serv11}.  Thanks to the excellent spatial resolution (half energy width = 0.8$\arcsec$), emission from the central SMBH in ESO 243-49 situated at just 8$\arcsec$ from HLX-1 is resolved. We extracted the photons in a 5$\arcsec$ radius centred on ESO 243-49.  We examined four different background regions of the same size as the source extraction region, placed above and below the galaxy bulge and to the left and right of the galaxy, as shown in \cite{serv11}, their figure 5. The background in each of these four regions was found to be the same.

\subsection{Swift data}

{\em Swift} XRT Photon Counting data dating from 2008 to 2016 (PI: O. Godet) were processed using the tool XRTPIPELINEv0.12.6, as described in \cite{webb12}.  The lightcurve of these data can be seen in Fig.~\ref{fig:XrayLC}. We analysed the spectrum of 11 ks of {\em Swift} data taken on the 13-14th September 2010. To do this we
took the grade 0-12 events, giving slightly higher effective area at
higher energies than the grade 0 events.  The
ancillary response files were created using XRTMKARF v0.5.9 and
exposure maps generated by XRTEXPOMAP v0.2.7. We fitted the
spectrum using XSPEC v12.7.0 with the response file
swxpc0to12s6\_20010101v013.rmf \citep{gode09b}.

\section{Results}

\subsection{HLX-1}

\begin{figure}
\includegraphics[width=9cm]{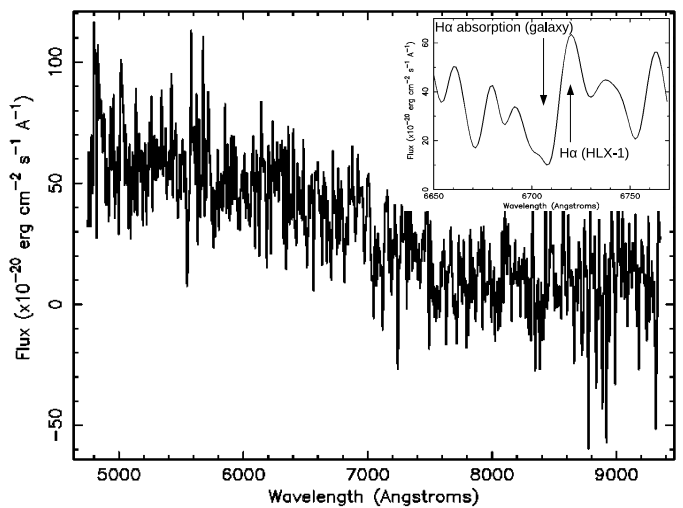}
\caption{Smoothed (Gaussian boxcar=3) galaxy subtracted {\em MUSE} spectrum of HLX-1. Insert shows the region around the H$_\alpha$ line.}
\label{fig:HLX1spectraMUSE}
\end{figure}

To investigate the nature of the H$_\alpha$ emission line associated with HLX-1 we extracted the HLX-1 spectrum using regions with a radius of 3 pixels from the {\em MUSE} datacube. To correct for the diffuse emission from the galaxy, background spectra were extracted using similar sized regions around HLX-1, where no point-like sources were detected and the emission scaled to equal that at the position of HLX-1. This was then subtracted from the HLX-1 spectrum.  The spectrum corrected for the galaxy emission can be seen in Fig.~\ref{fig:HLX1spectraMUSE}. The blue nature of the spectrum, as noted in \cite{farr12} is immediately obvious. Estimating the (AB) V and the R band magnitude from the spectral continuum we find values of V=24.97$\pm$0.25 and R=24.76$\pm$0.25 (1 $\sigma$) and comparing these to previous values \citep[e.g.][]{webb14} we find that the magnitudes are comparable to the faintest detections that were made during the low/hard state.

We investigated the region around 6721 \AA\ (see insert of Fig.~\ref{fig:HLX1spectraMUSE}) where \cite{wier10} and \cite{sori13} detected an emission line with a flux $\sim$1-2 $\times$ 10$^{-17}$ erg cm$^{-2}$ s$^{-1}$. Such a line, if the flux had not varied since the previous detections using {\em FORS2} on the {\em VLT} (which were made during the high/soft state, see Fig.~\ref{fig:XrayLC}), should be detected with a signal to noise of 4.7 per spectral bin of 1.1\AA. No significant emission line was detected at this wavelength, although a fluctuation, resembling an emission line, with a signal to noise ratio (SNR) of $\sim$2 (signal to noise of the continuum $\sim$0.8, 1$\sigma$ detection) was noted (see inset of Fig.~\ref{fig:HLX1spectraMUSE}). This fluctuation is centred at 6720.3 \AA, and has a flux of $\sim$3 $\times$ 10$^{-18}$ erg s$^{-1}$ and a FWHM$\sim$12.5 \AA, similar to the value found by \cite{sori13}.  We take the flux value as the 1 $\sigma$ upper limit for the emission line, indicating that the line flux has varied by a factor of $>$6. 

We also searched for other emission lines that may also be observed from the region around an ultra-luminous X-ray source or an IMBH. We searched for H$_\beta$ (4861 \AA), [O {\sc III}] (4959 and 5007 \AA) as well as various H{\sc I} lines. However, we found no evidence for any of these lines, where  the 1 $\sigma$ upper limits on these lines are 1.2$\times$10$^{-18}$ erg cm$^{-2}$ s$^{-1}$ for H$_\beta$, 1.6$\times$10$^{-19}$ erg cm$^{-2}$ s$^{-1}$ for [O {\sc III}] (4959 \AA) and 4.7$\times$10$^{-19}$ erg cm$^{-2}$ s$^{-1}$ for the 5007~\AA\ line.

The {\em X-shooter} data were also taken during the low hard state. We examined the HLX-1 spectrum corrected for the galaxy emission. No emission lines around 6721 \AA\ were found (both the gasgano reduced and the ESO reduced spectra were analysed). If the flux was the same as that observed during the high state, we would expect to detect the line with signal to noise of $\sim$3 per spectral bin of 0.3 \AA.  However, if the flux in the line was similar to the upper limit identified from the {\em MUSE} data, then a signal to noise of $\sim$0.9 per spectral bin of 0.3\AA\ would be expected, insufficient to detect the line.  The flux level in the continuum for HLX-1 in the low hard state is expected to be negligible (signal to noise of 0.05-1.0 per pixel in the UV and VIS wavelength ranges using the {\em X-Shooter} exposure time calculator), making it difficult to place upper limits on other possible emission lines. None of the lines investigated in the {\em MUSE} data, nor O {\sc II} (3726 and 3729 \AA), or the Bowen blend (4686 \AA) nor the Paschen and Brackett lines, were detected.

\subsection{The galaxy ESO 243-49}

\subsubsection{The mass of the central supermassive black hole}
\label{sec:MassSMBH}

\begin{figure}
 \centering
\includegraphics[width=9cm]{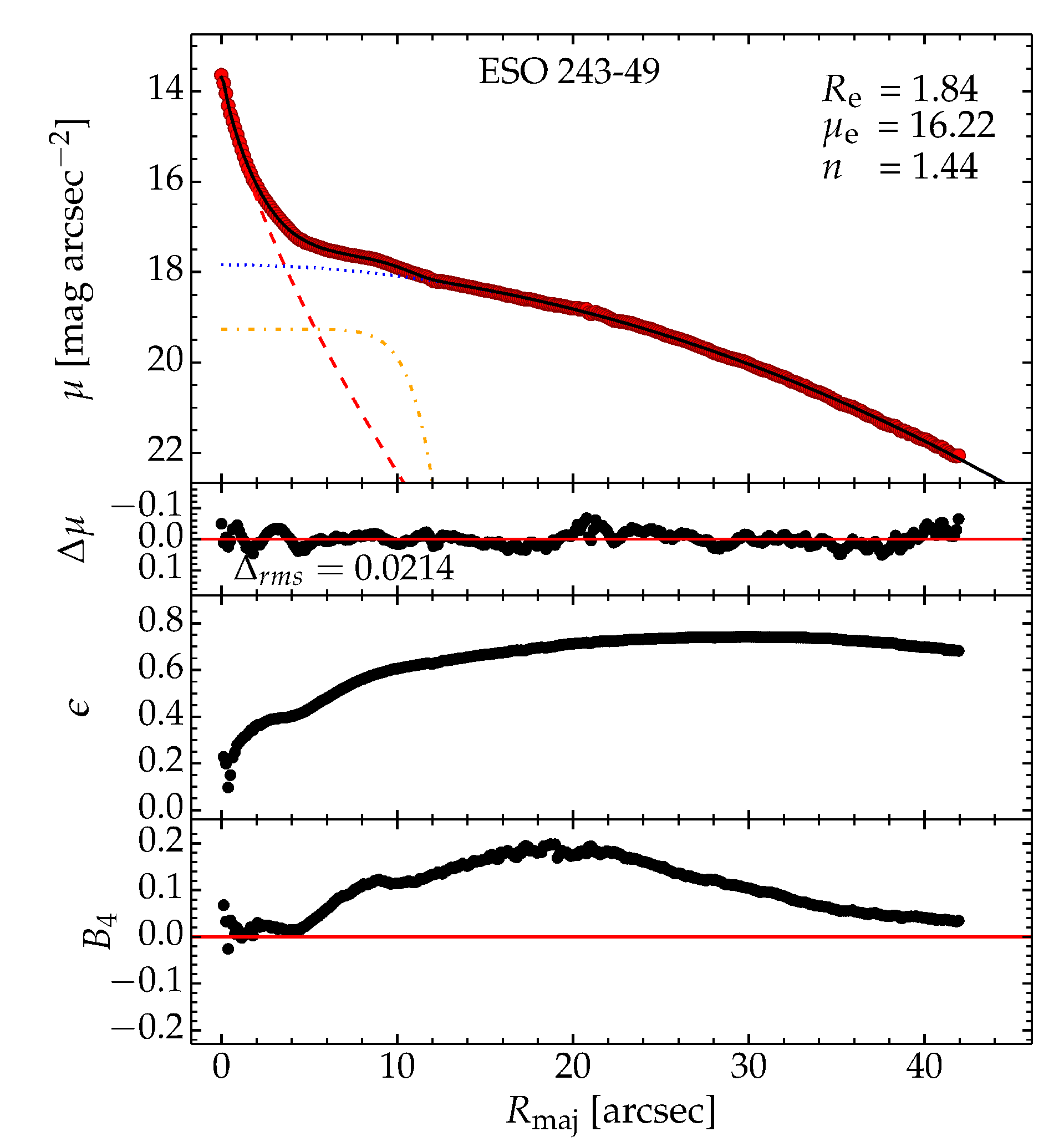}
\caption{$H$-band major-axis surface brightness profile of ESO~243-49. The red circles are the data extracted from the $HST$ image, while the black curve represents the best-fitting model, which in turn is built up of a spheroid/bulge (red dashed line), a bar (yellow dot-dashed line) and an edge-on disc (blue dotted line). The inset numbers are the S\'ersic parameters of the bulge. The residual profile is shown in the next panel below, followed by the ellipticity and $B_4$ amplitude (discyness) profiles.}
\label{fig:decompositionESO243}
\end{figure}

Figure~\ref{fig:decompositionESO243} displays, from top to bottom: the modelled {\it HST} major-axis light profile, the residual profile ($\Delta\mu$), the isophote ellipticity ($\epsilon$) profile (also Fig.~\ref{fig:decompositionESO243}) and finally the $B_4$ harmonic amplitude (the measure of the isophote's boxiness or discyness) profile.  In addition to the bulge and edge-on disc components, the light profile displays the characteristic signature of a bar. We determined a major-axis bulge S\'ersic index \citep{grah05} of $n = 1.44 \pm 0.08$, and half-light (effective) radius of $R_{\rm e} = 1.84 \pm 0.11\arcsec$.  The (black hole mass)--(host spheroid S\'ersic index) scaling relation of \cite{savo16} yields a SMBH mass estimate of $1.4 \times 10^7 M_{\odot}$, with a $1\sigma$ interval of $(0.7 - 3.0) \times 10^7 M_{\odot}$.

To compute the total magnitude of each component, we mapped the major-axis profile onto the `equivalent-axis', which circularises each isophote such that its enclosed area is conserved ($R_{\rm eq} = R_{\rm maj} \sqrt{1-\epsilon(R_{\rm maj})}$, where $\epsilon$ is the isophote ellipticity). We determined the $H-$band bulge apparent magnitude to be $m_H = 12.72\pm0.10$ [Vega] mag, and a bulge-to-total flux ratio of $0.25\pm0.06$, making this a barred S0 or S0/a galaxy. The mass of the central SMBH in ESO~243-49 was estimated using the (black hole mass)--(host spheroid luminosity) relation \citep{savo15} to be $2.3 \times 10^7 M_{\odot}$. Taking into account the scatter on the relation places the mass in the ($1\sigma$) range of $(0.5 - 11) \times 10^{7} M_{\odot}$. As this scaling relation is calibrated for 3.6 $\mu$m, we assumed a $(H - 3.6 \mu m)$ colour of $0.55\pm0.18$ \citep{savo15}.


We extracted the {\em MUSE} spectra from within the effective radius determined from the HST data and determine its velocity dispersion using the Penalized PiXel-Fitting software \citep[pPXF][]{capp16,capp04}. The PSF was determined to be less than 1$\arcsec$, much smaller than R$_e$, therefore we did not correct our measurement for the PSF. To determine the central velocity dispersion, we used the full MILES stellar library \citep{sanc06,falc11} and the same method as described in Section~\ref{sec:kinematics}. This gives a central velocity dispersion of 224$\pm$1 km s$^{-1}$, which is fairly high, but may be due to the bar discerned in this galaxy \citep[e.g.][]{hart14}. Using the relationship for a barred S\'ersic galaxy given in \cite{grah13} we determine a black hole mass of 10.6 $\times$ 10$^7$ M$_\odot$. Taking into account the RMS scatter on the relation gives a mass range of 4.8 -- 23 $\times$ 10$^7$ M$_\odot$.

The X-ray photons extracted from the region surrounding the SMBH in ESO 243-49  in the {\em Chandra} data, are soft. Fitting a typical galaxy spectrum to these photons (mekal in $xspec$), as described in \cite{serv11}, we determine an unabsorbed luminosity in the 0.2-10 keV domain of $\sim$6 $\times$ 10$^{39}$ erg s$^{-1}$, assuming a source
distance of 95 Mpc \citep[using the WMAP cosmology,][]{benn13}. The ATCA data reveal a 5 GHz flux of 180$\pm$15 $\mu$Jy/beam (1 $\sigma$ error). To estimate the mass of the black hole using the X-ray and radio fluxes/luminosities, we used the sample presented by \cite{merl03}, which includes black holes in all states and the correlation presented from the analysis of this sample by \cite{koer06}. The energy bands used in \cite{koer06} are better suited to the ESO 243-49 SMBH,  as the X-ray emission is soft and shows very few counts in the 2.0-10.0 keV band used by \cite{merl03}, leading to large uncertainties in the X-ray luminosity.  We determine a mass of 2.9 $\times$ 10$^8$ M$_\odot$. Taking into account the errors on the radio and X-ray fluxes, as well as on the relationship gives a mass range of  0.029 -- 290 $\times$ 10$^8$ M$_\odot$.  However, \cite{serv11} state that the X-ray emission appears to be extended and therefore it may not come from the central SMBH, but from diffuse gas around the central black hole, in which case this mass estimate is not reliable.

Although the radio and X-ray data are taken a week apart, analysing the {\em Swift} data taken on the same day as the radio data, we do not see any strong evolution of the flux from the SMBH. We prefer, however, to use the luminosity estimated from the {\em Chandra} data, as this is more accurate thanks to a better spatial resolution than {\em Swift}.

\subsubsection{Stellar and gas kinematics}
\label{sec:kinematics}

\begin{figure*}
 \centering
\includegraphics[width=6cm]{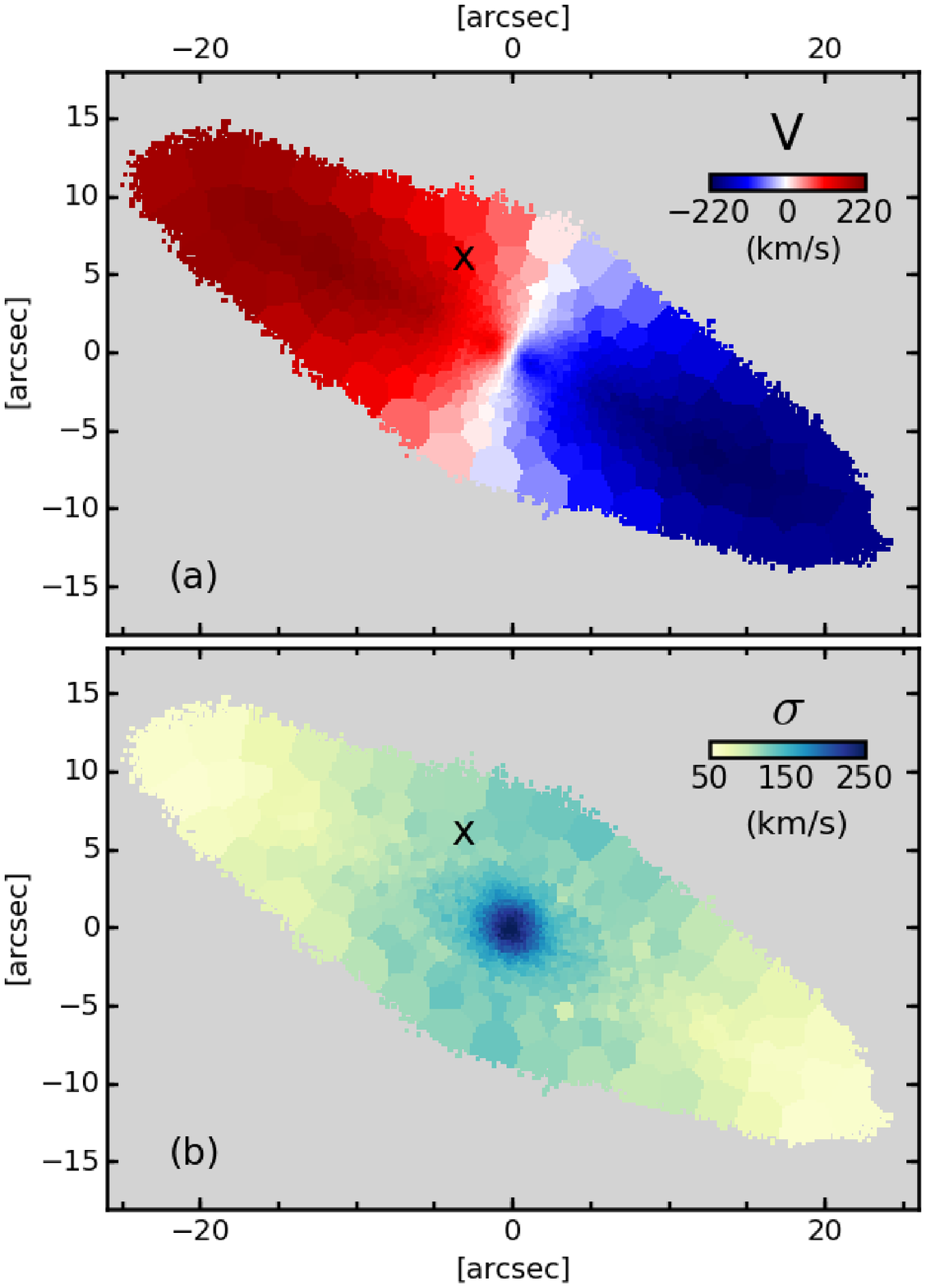}
\includegraphics[width=6cm]{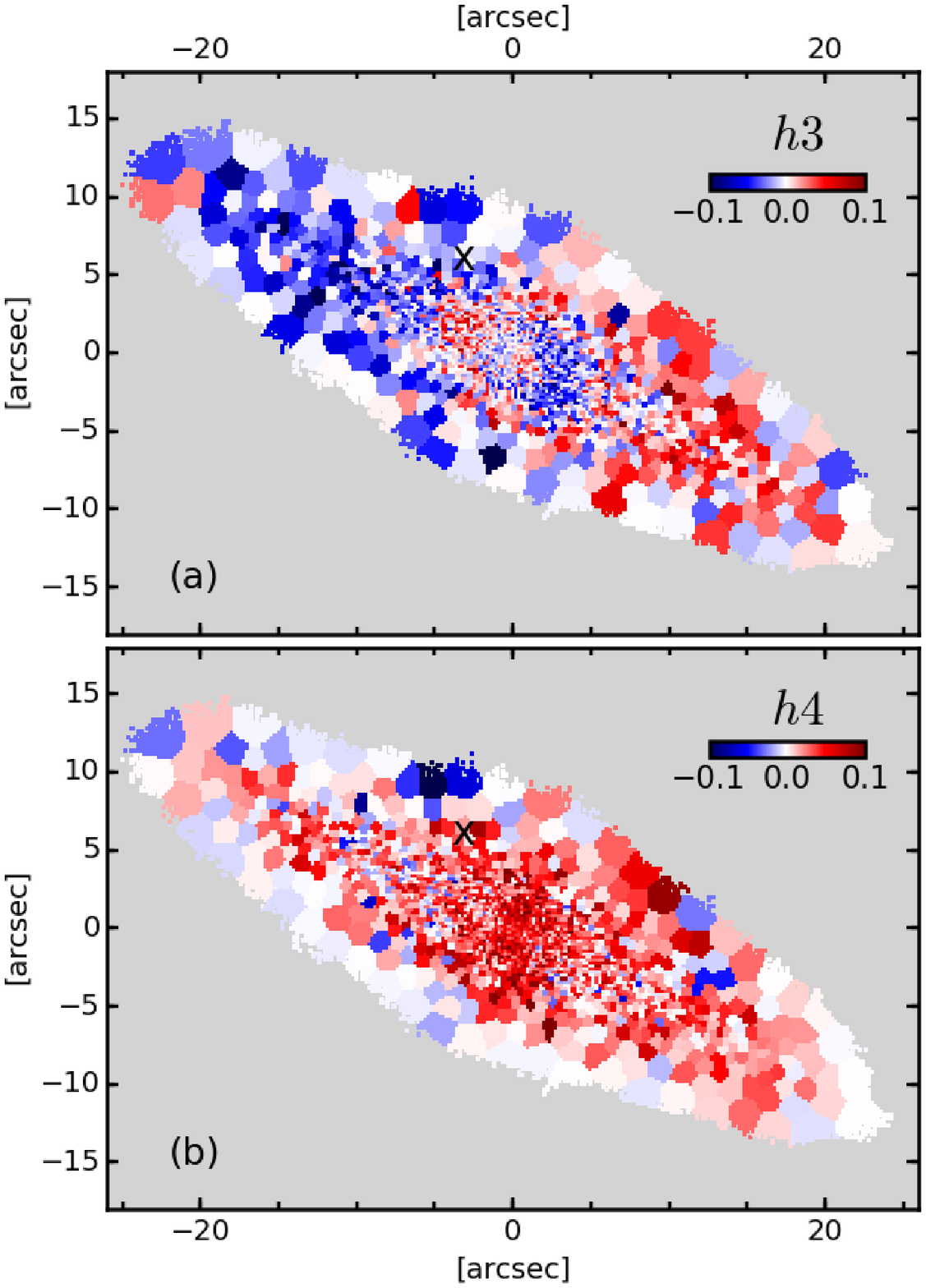}
\includegraphics[width=6cm]{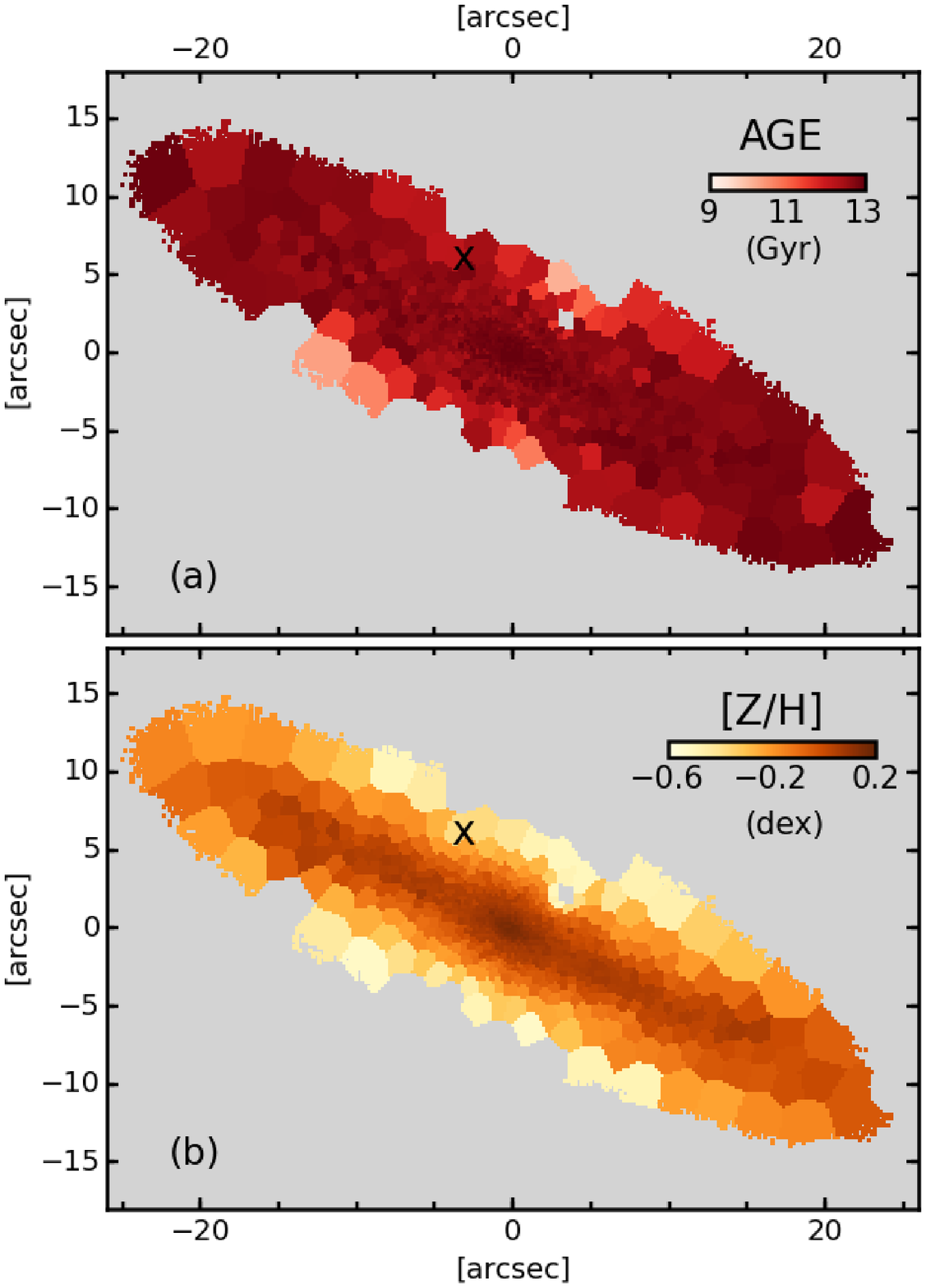}
\caption{Top left: Stellar velocity in ESO 243-49 as determined from the {\em MUSE} data. Bottom left: Stellar velocity dispersion in ESO 243-49. Top middle: Skewness of ESO 243-49. Bottom middle: Kurtosis of ESO 243-49. Top right: Age. Bottom right: Metallicity in ESO 243-49. Each figure is centred on ESO 243-49 and the distance in arcsec is the offset from the galaxy centre in right ascension and declination.  The X marks the position of HLX-1.}
\label{fig:ESO243}
\end{figure*}

We also investigated the kinematics of the galaxy \object{ESO 243-49}. To do this we binned the combined data cube in the spatial directions to create pixels (spaxels) with a signal to noise of 50. We chose a signal to noise of 50 as it allows  robust extraction of the kinematics and stellar populations while keeping a reasonable spatial sampling \citep[e.g.][]{capp11}. We used the adaptive spatial binning software developed by \cite{capp03}, based on Voronoi tessellation. We estimated the original SNR of each spectrum by taking the ratio of the signal and the square root of the variance produced by the {\em MUSE} pipeline. We took the median of that ratio between 5500 \AA\ and 6000 \AA\ which is representative of the wavelength range used for extracting the stellar kinematics and population properties. The adaptive spatial binning delivered more than 62\% of the bins with a SNR equal to or greater than 50. The central 3\arcsec\ were not binned as the signal to noise was above 50 for each pixel, with a scatter of 6 and a minimum signal to noise of 40. 

We used the full MILES stellar library that covers the wavelength range 3525 - 7500 \AA\ with a constant spectral resolution of 2.50 \AA\ FWHM, similar to but lower than the {\em MUSE} spectral resolution. Since the difference in spectral resolution between the {\em MUSE} data and the MILES stellar library, i.e., 0.2 \AA\ FWHM at 5000 \AA\ ($\sigma$ = 5 km s$^{-1}$), is significantly lower than ESO 243-49 velocity dispersion ($\sigma >$ 50 km s$^{-1}$), we did not convolve the stellar library to the same spectral resolution as the observed spectra. We derived the first four order moments of the line of sight velocity distribution (LOSVD), namely stellar velocity (V), velocity dispersion ($\sigma$), skewness of the stellar distribution ($h_3$) and kurtosis of the stellar distribution ($h_4$), for each spaxel of the {\em MUSE} cube, in the following manner. We set up pPXF to use additive polynomials of the 8th order and the default value of penalisation (0.4). We restricted the fitted wavelength range to 4600 - 6800 \AA\ as it contains strong absorption lines (H$_\beta$, Fe, etc.) as well as to avoid spectral regions highly contaminated by sky lines residuals ($\lambda >$ 7000 \AA). We masked five bright sky emission lines at 5577 \AA; 5889 \AA; 6157 \AA; 6300 \AA, and 6363 \AA, as well as the potential H$_\alpha$ and N {\sc II} gas emission lines.  We first fitted the stacked spectrum of the central 2\arcsec with the full MILES stellar library, and then used the best fit (resulting from a combination of $\sim$30 original MILES templates) as the stellar template to fit each spatially binned spaxel of the {\em MUSE} data.

\begin{figure}
\includegraphics[width=9cm]{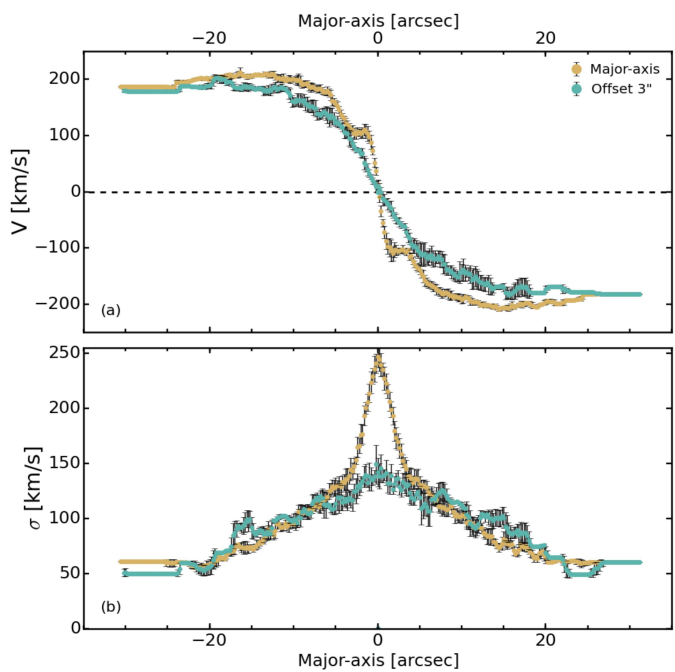}
\caption{Top: The stellar velocities (km s$^{-1}$) as a function of the distance from the centre of ESO 243-49.  Bottom: The stellar velocity dispersion (km s$^{-1}$) as a function of the distance from the centre of ESO 243-49.  Orange points are along the major axis and blue points are offset at 3\arcsec.}
\label{fig:Vcut}
\end{figure}

The galaxy is clearly rotating with speeds up to $\sim$200 km s$^{-1}$ (see Fig.~\ref{fig:ESO243}, top left). We discern a disc and an inner thin disc in Fig.~\ref{fig:Vcut} as shown in \cite{come16}. There is also evidence for the presence of a slower (than the disk) spheroid component, see Fig.~\ref{fig:ESO243} (top middle) in which an anti-correlation of the skewness parameter can be seen with respect to the velocity field in Fig.~\ref{fig:ESO243}, \citep[top left, e.g.][]{bend94,kraj08}. The correlation of the skewness parameter with the velocity field along the major axis around 3-5\arcsec\ is likely to be associated with the inner-disk, as evidenced from modelling the H-band photometric data of the bulge of ESO 243-49, see Section~\ref{sec:MassSMBH}. We extracted optical spectra of the extremities of this disc, from the {\em MUSE} data. Strong H$_\alpha$ and [N {\sc II}] (6583 \AA) emission lines are detected in this region. We used the Doppler shift of these lines to determine the rotation rate of the gas in the disc. The H$_\alpha$ line and the [N {\sc II}] are centred at wavelengths of 6717.4$\pm$0.5 \AA\ and 6738.4$\pm$0.5 \AA\ on the left side of the disc respectively and at 6703.7$\pm$0.5 \AA\ and 6725.3$\pm$0.5 \AA\ on the right side of the disc respectively. This indicates maximal speeds of 300 km s$^{-1}$. We find that the radius of the inner disc is 3.9$\arcsec$ or 1.8 kpc. 

Using the spectrum extracted from the central pixel of the galaxy we determine the wavelength of the  H$_\alpha$ and [N {\sc II}] (6583 \AA) emission lines, as well as the Na D (5892.5 \AA), Mg {\sc I} (5175.4 \AA) and Ca (5269.0 \AA) absorption lines due to the galaxy. We detect these lines at 6710.6 \AA, 6731.0\AA, 6025.5\AA, 5290.7\AA\ and 5387.3\AA\ respectively which gives a redshift of the galaxy of $z=0.0225\pm0.0001$. Searching for any perturbation to the galaxy that could be caused by a merger event, we examine the stellar velocity distribution, the stellar velocity dispersion, the stellar skewness and kurtosis of ESO 243-49 as well as the stellar age and metallicity distributions (see Figs.~\ref{fig:ESO243} and \ref{fig:Vcut}). All of the distributions are smooth, with no region showing different behaviour from the rest of the galaxy. The rotation of the galaxy is homogeneous as is the velocity dispersion. The exception is the more rapidly rotating disc in the centre of the galaxy seen in the skewness distribution (i.e., central anti-correlation seen in the skewness (h3) and stellar velocity (V) maps). The stellar velocity dispersion peaks in the central regions of the galaxy (following the high velocity gradient in this region) and falls off more slowly to the exterior where the velocity gradient is low, as would be expected for a non-perturbed galaxy (see also Fig.~ \ref{fig:Vcut}). A similar result has also been seen by \cite{come16}.

We also examine the age and metallicity of the stellar population of \object{ESO 243-49} (see again Fig.~\ref{fig:ESO243}), using the MIUSCAT library \citep{vazd12}. We used a subset of 120 models from the MIUSCAT library covering an age range of 0.5 to 14.1 Gyr, and a metallicity range of -0.71 to +0.22 dex, therefore all included in the "SAFE" range of the library \citep[see][for more details]{vazd12}. These models were computed using a unimodal IMF with a slope of 1.3 and based on Padova isochrones \citep{gira00}. We used the same pPXF settings as for the kinematics analysis except that we used only multiplicative polynomials of the 12th order. The age and metallicity values are mass-weighted and obtained from a regularisation approach as in \cite{guer16}. Some spatial bins gave very young ages but the spectra showed relatively strong sky residuals which biased the fits so these bins were not deemed reliable and were masked in the plots. These masked bins are located in the outskirts of the galaxy.  This analysis reveals that the central disc is old ($\sim$12-13 Gyr) and metal-rich \citep[see Fig.~\ref{fig:ESO243}, top and bottom right and consistent with the independent analysis of][see their figure 5]{come16}.  There is nothing peculiar in other regions of the galaxy nor at the position of \object{HLX-1}.

\subsubsection{Other objects in the ESO 243-49 field of view}
\label{sec:background_sources}

To better understand the dynamics in \object{ESO 243-49}, we extracted the {\em MUSE} spectra of the brighter globular clusters, identified using the HST observations. However, due to their faint nature, no useful spectral information could be obtained.

Examining the spectra of other sources in and around \object{ESO 243-49} reveals some interesting objects. Firstly, the {\em blue patch} noted in \cite{wier10} and \cite{sori13} located at 01$^h$10$^m$27$\fs$99 -46$^\circ$04$^\prime$23$\farcs$55 and with extent of $\sim$11 pixels ($\sim$2$\farcs$2) shows a very strong emission line at 6763.3\AA\ (wavelength in air), with an equivalent width of -125 \AA, likely to be H$_\alpha$ at a redshift of $z=0.0306\pm0.0001$.  A second line at 5010.0 \AA\ with an equivalent width of -350 \AA\ is then H$_\beta$ and we also identify O{\sc III} (5007\AA) at 5160.0 \AA\ and two other significant lines at 5576.5 \AA\ and 6088.87 \AA.  The extent, spectral features and redshift of this source implies that it is a background galaxy.  Four other background objects are also identified thanks to their H$_\alpha$ and H$_\beta$ lines, but at much greater redshifts. The positions and redshifts of these sources can be seen in Table~\ref{tab:bkg_objects}.

\begin{table}
\caption{Background objects in the field of view of ESO 243-49}\label{tab:bkg_objects}
\centering
\begin{tabular}{ccc}
\hline \hline
RA & dec. & z  \\
\hline
01$^h$10$^m$27$\fs$99 & -46$^\circ$04$^\prime$23$\farcs$55 & 0.0306$\pm$0.0001\\
01$^h$10$^m$27$\fs$07 & -46$^\circ$04$^\prime$32$\farcs$55 & 1.280$\pm$0.004 \\
01$^h$10$^m$28$\fs$55 & -46$^\circ$03$^\prime$56$\farcs$15 & 1.13$\pm$0.01 \\
01$^h$10$^m$29$\fs$74 & -46$^\circ$04$^\prime$47$\farcs$95 & 1.140$\pm$0.002 \\
01$^h$10$^m$27$\fs$76 & -46$^\circ$04$^\prime$52$\farcs$35 & 1.240$\pm$0.003 \\
\hline
\end{tabular}
\end{table}

\section{Discussion}

The non-detection of the H$_\alpha$ line in the vicinity of HLX-1 in the {\em MUSE} and the {\em X-Shooter} data indicates that the  H$_\alpha$ line flux has diminished by more than a factor 6 since HLX-1 was observed in the high/soft state.  The line flux would not be expected to vary by at least a factor 6 over a year if the emission were to emanate from a nebula, as the size of such a nebula is expected to be much larger than the distance travelled by the electromagnetic radiation from the central X-ray source over one year. The variability of the H$_\alpha$ line also implies that the emission is not related to star formation, as no such variations on this timescale are expected. The variability suggests that the H$_\alpha$ emission comes from close to the black hole, either from the disc or possibly from a corona, thus confirming the distance of $\sim$95 Mpc and re-enforcing its association with the galaxy \object{ESO 243-49}.  Further, the fact that no [O {\sc III}] (5007 \AA) is detected in the {\em MUSE} or the {\em X-Shooter} spectra, nor He II (4686 \AA) in the {\em X-Shooter} spectrum, counts against a ULX bubble \citep{sori13}. 
 
Confirming the distance to \object{HLX-1} also corroborates that \object{HLX-1} shows sub-Eddington outbursts \citep[e.g.][]{gode14}. \cite{laso15,laso11} state that if HLX-1 is indeed at 95 Mpc and shows sub-Eddington outbursts, then the outbursts can not be driven by the thermal–viscous disc instability. The outbursts are then likely to be triggered as material from a companion star in a highly eccentric orbit, impacts the inner regions of the accretion disc around the IMBH, as the star passes at periastron \citep{webb14}. In this case the mass-transfer propagation cannot be viscous but must be mediated by waves, as proposed in \cite{webb14,laso11}. We have demonstrated that such a binary could form and remain bound for a number of orbits \citep{gode14}. \cite{gode14} demonstrated, however, that such a system would then slowly become unbound, due to tidal effects and increasing mass transfer. This indeed appears to be occurring, if the outbursts we observe are due to the passage at periastron of the companion star. We have observed four full orbits that last approximately one year and since then, the subsequent orbits have been 13.5 months, 15 months and $>$26 months (see Fig.~\ref{fig:XrayLC}). The alternative to this scenario, as suggested by \cite{king14}, proposes that the variability seen is due to the precessing of relativistic jets. However, this is ruled out by the fact that if the increase in luminosity is due to the jet precessing into our line of sight, we would expect the X-ray emission to become harder due to the Doppler boosting, but instead we observe softer emission \citep{gode09a,serv11}. Further, as the H$_\alpha$ emission can not be due to star formation, as noted above, then the H$_\alpha$ may be produced by photo-ionisation by the X-ray source or collisional ionisation by an outflow.  The H$_\alpha$ emission should be nearly isotropic for any situation where it is powered by photonisation or a low speed shock.  The relatively narrow line suggests that this is the case for HLX-1.  As a result, if it is powered by photoionisation, it should be a tracer of the luminosity of the source averaged over $4\pi$, and if it is powered by a shock, it should be constant in luminosity.  The fact that the H$_\alpha$ emission is detected from this source near its peak, despite the low column density local to the source, suggests that there is not a strong beaming of the X-ray flux \citep[see also][]{farr11}.  The fact that the luminosity varies, such that it is found to be a factor of at least 6 fainter when the X-ray luminosity is low than near the X-ray peak, indicates that the emission is not from collisional ionisation in a large shock.  Therefore, the totality of the evidence from the H$\alpha$ emission indicates that the X-ray source is not strongly beamed.  These constraints apply generically to beaming models, and not only to the particular model of \cite{king14}.

As explained in Sec.~\ref{sec:intro}, HLX-1's behaviour shows many similarities with black hole X-ray binaries. Indeed black hole X-ray binaries often show the H$_\alpha$ line luminosity dropping as the X-ray and optical emission drops \citep{fend09}. \cite{hyne02} also showed that the  H$_\alpha$ emission was correlated with the continuum emission in V 404 Cyg and \cite{jin12} and \cite{pane06} showed a correlation between the H$_\alpha$ emission and the X-ray luminosity in active galactic nucleii (AGN). Further, \cite{dewa08} showed the same correlation for AGN with lower mass black holes, approaching the mass of the black hole in HLX-1, supporting the idea that we could expect to observe less H$_\alpha$ flux in the low/hard state than was observed in the high/soft state.

Alternatively, the variability may not be linked to the X-ray state, but the phase at which we observed HLX-1, in a similar way to that observed in the black hole binary XTE J1118+480 \citep{torr04}.  If we suppose that HLX-1 goes into outburst as the companion star reaches periastron \citep{laso11}, which we can call phase 0, and that the time between outbursts is the time taken for the companion to orbit the black hole, the spectroscopy presented by \cite{sori13} was taken shortly after this time (roughly at phase 0) and that presented in \cite{wier10} was taken almost three months after phase 0 (roughly at phase 0.25). The {\em X-Shooter} data were taken between one month and one week before outburst, i.e. around phase 0.95 given the roughly year long orbit. The {\em MUSE} data were taken 4 months before outburst (where that orbit had a 15 month duration), i.e. around phase 0.75. The data taken at different points in the orbit may then lead to the variability observed.


The four mass measurements of the central SMBH in ESO 243-49 are very similar. The estimate using the black hole fundamental plane has larger error bars than the other estimates. It should also be noted that the X-ray emission appears to come from hot diffuse gas \citep{serv11} and not from an accretion disc. Further, there is no evidence for a low luminosity AGN in the optical spectra. The X-ray emission may then be due to star formation in the central region, which may also account for the radio emission. Therefore the mass estimate using this method should be used with caution and at best as an upper limit. We therefore take only the mass estimates from the two other methods which give a 1$\sigma$ range of  0.5-23 $\times$ 10$^7$ M$_\odot$  for the mass of the central SMBH in ESO 243-49.

The rapidly spinning disc discerned in the {\em MUSE} data, compared to the slower bulge is compatible with mergers in the history of the galaxy, e.g. \cite{arno14}.  \cite{mape15} found using Smoothed Particle Hydrodynamic simulations, that the fingerprints of minor mergers of dwarf galaxies with elliptical galaxies is the formation of a gas ring in the S0 galaxy. This ring may be related to the orbit of the satellite galaxy, and its lifetime should depend on the merger properties.  The high gas velocity observed from the inner disc may be related to the bar discerned in Sect.~\ref{sec:MassSMBH}, which could be oriented towards the observer (boxy bulge, but not peanut) and would then explain the high gas velocity. Indeed the isophotal fitting supports this interpretation.

Whilst the current {\em MUSE} data is insufficient to examine the region around HLX-1 and determine whether this object is associated with ESO 243-49 due to a minor merger, as first proposed in \cite{webb10}, or whether it was born in situ, the rapidly rotating central disc in ESO 243-49 is compatible with the idea that such mergers may have taken place during the history of the galaxy.  Given the close proximity of ESO 243-49 to the most massive central galaxy in the cluster, a number of close satellites and minor mergers are to be expected, as noted in Sec.~\ref{sec:intro}.   If there has been a dry minor merger in the past, it should have taken place more than a dynamical time ago, i.e.  t $>$ 300 Myr as we no longer see any evidence of such an event \citep{barn92,duc13}. However, with regards to a minor-merger that gave rise to HLX-1, it is less clear. Maybe the minor-merger is still ongoing, with the orbit of the dwarf galaxy being highly eccentric so that the timeframe for the dwarf galaxy to encounter the host galaxy again would be long. Without additional radial velocity for the system, it is difficult to assess the status of the merger and only with future data and detailed computer modelling will it be possible to estimate the time of the latest minor-merger.


\begin{acknowledgements}
Based on observations made with ESO Telescopes at the La Silla Paranal Observatory under programme IDs 60.A-9328 and 091.D-0823. We thank the referee for constructive and pertinent suggestions that helped to improve this paper.  NW, MC, OG, DB, AG, AD, and MM acknowledge the Centre National d’Etudes Spatiales (CNES) for their support.
\end{acknowledgements}

%
   \bibliographystyle{aa} 
   \bibliography{ESO243v6} 
%

\end{document}